# The role of local-geometrical-orders on the growth of dynamic-length-scales in glass-forming liquids

Kaikin Wong[1], Rithin P. Krishnan[1], Changjiu Chen[1], Qing Du[2], Dehong Yu[3], Zhaoping Lu[2], and Suresh M. Chathoth[1*]

[1]Department of Physics and Materials Science, City University of Hong Kong, Kowloon Tong, Hong Kong, P. R. China.

[2]State Key Laboratory for Advanced Metals and Materials, University of Science and Technology, Beijing, 100083, P. R. China.

[3]Australian Nuclear Science and Technology Organization, Lucas Height, 2234, Australia.

## Abstract

The precise nature of complex structural relaxation as well as an explanation for the precipitous growth of relaxation time in cooling glass-forming liquids are essential to the understanding of vitrification of liquids. The dramatic increase of relaxation time is believed to be caused by the growth of one or more correlation lengths, which has received much attention recently. Here, we report a direct link between the growth of a specific local-geometrical-order and an increase of dynamic-length-scale as the atomic dynamics in metallic glass-forming liquids slow down. Although several types of local geometrical-orders are present in these metallic liquids, the growth of icosahedral ordering is found to be directly related to the increase of the dynamic-length-scale. This finding suggests an intriguing scenario that the transient icosahedral ordering could be the origin of the dynamic-length-scale in metallic glass-forming liquids.

After many decades of extensive studies, there is still no generally accepted and fundamentally understood reason for the dramatic growth of structural relaxation time when glass-forming liquids cool towards the glass-transition temperature[1,2]. Numerous experiments and molecular dynamics simulation results suggest that the occurrence of cooperative process in glass-forming liquids; this in turn implies the presence of cooperative rearranging regions (CRR) over a length scale. The CRR was first invoked by Adam and Gibbs in their description of the dynamics of glass-forming liquids[3]. The Adam-Gibbs theory connects relaxation time with growing length scales and thermodynamics via configurational entropy. The configurational entropy implies the atomic ordering over certain length scales in glass-forming liquids can be of static and dynamic nature. Recent experimental and theoretical research has been focused on growing static and dynamic length scales when glass-forming liquids cooled towards the glass-transition temperature[4–6].

A variety of experiments and numerical studies have shown that the local dynamics in glass-forming liquids is spatially heterogeneous[7–10]. Dynamic heterogeneity (DH) is the term used to describe this aspect, and the presence of DH in liquids has been numerically characterised by studying the non-Gaussian distribution of the displacement of individual particles during a fixed time interval[11]. Experiments and numerical results have shown that particles with similar mobilities tend to form clusters, which resulted in the existence of spatial correlations of local mobility[12,13]. A well-defined clusters of cooperatively moving localised fast particles have been seen in colloids near the glass-transition[12]. By simulation, clusters of cooperatively moving particles with string-like structures have been identified[13]. These observations clearly indicate that the one or more length-scales, which characterizes the spatial correlation of local mobility, can be obtained by analysis of the size distribution of these clusters. In practice, it is difficult to extract a quantitative estimation of the lengthscale since the definition of slow and fast is somewhat arbitrary[14]. Consequently, the quantitative information about the length scale of DH has been obtained from multi-point correlation functions[14].

On the other hand, the static structure factor measured at different temperatures of a glass-forming liquid does not indicate any obvious structural change, although the dynamics can be several orders of magnitude different[15,16]. To the contrary, the static length scales in glass-forming liquids computed by different methods, such as point-to-set correlation[17], configurational entropy[18], and bond-orientational order[19] demonstrate that it is of the order of interparticle distance and grows by a factor of 2 to 9 as the glass-forming liquids cool towards a mode-coupling critical temperature, $T_c$ [18,20,21]. However, the static length scale increases at a slower pace than the dynamic-length-scale in decrasing the temperature of glass-forming liquids. This indicates that the growth of these length scales are not correlated. The growth of local geometrical-orders (LGOs) is also argued to be the cause of the rapid rise in the relaxation time of cooling glass-forming liquids[22–24]. In three dimensional systems of monodispersed hard-spheres, the icosahedron is the most locally-preferred structure and increasing its number while cooling is believed to be linked directly to vitrification[25,26]. In multicomponent, polydispersed metallic systems, an increasing number of five-fold symmetry clusters is reported to be the reason for their better glass-forming ability (GFA)[27]. However, the role of specific LGO on the growth of dynamic-length-scale is not yet undersood, although numerous studies have shown rapid growth of dynamic-length-scale and LGOs on cooling glass-forming liquids towards the glass-transition temperature $T_g$. In this article, we studied a ternary metallic glass-forming system to understand whether the growth of any specific LGO correlates with the increase of dynamic-length-scales by using quasielastic neutron scattering (QENS) and molecular dynamics (MD) simulation techniques.

The Cu-Zr-Al is a well-known glass-forming metallic system with a distinct GFA[28]. For this study, we have chosen the following composition: $(Cu_{50}Zr_{50})_{100-x}Al_x$ ($x$ = 2, 4, 8, and 10). The GFA of the system is found to increase with the addition of Al in $Cu_{50}Zr_{50}$[28]. The QENS experiments were conducted on a time-of-flight neutron scattering instrument, Pelican, at Bragg Institute in Sydney, Australia[29] (see Supporting Information). The self-intermediate scattering functions, $\Phi(Q,t)$s obtained for the $(Cu_{50}Zr_{50})_{96}Al_4$ liquid at different temperatures are shown in Fig. 1a. The $\Phi(Q,t)$ data were fitted with the Kohlrausch-William-Watt (KWW) function $\left[\Phi(q,t) = f_q \exp-\left(\frac{t}{\tau_\alpha}\right)^\beta\right]$, where $f_q$ is the Debye-Waller factor, $\tau_\alpha$ is the relaxation time, and $\beta$ is the stretching exponent. The value of the stretching exponent was found to be $\beta < 1$ and temperature dependent, but composition- and Q-independent. At the lowest measured temperature, we obtained a value of $\beta = 0.6 \pm 01$, and at the highest temperature $\beta = 0.8 \pm 0.1$.

The stretching of $\Phi(q,t)$ indicates that multiple relaxation processes in alloy liquids, even at temperatures well above the melting temperature, and the presence of heterogeneous dynamics in CuZrAl liquids. As we mentioned earlier, the DH can be estimated by calculating the four-point correlation function $\chi_4(t)$, but this quantity cannot be evaluated from a QENS experiment. However, recent theoretical advances have shown that the $\chi_4(t)$ is related to the dynamic susceptibility, $\chi_T(Q,t)$ by the fluctuation-dissipation theorem[30]. The $\chi_T(Q,t)$ can be evaluated from the $\Phi(Q,t)$ which is readily obtained from QENS experiments. The dynamic susceptibilities were obtained by $\chi_T(Q,t) = \frac{\partial \Phi(Q,t)}{\partial T}$. Fig. 1b shows the $\chi_T(Q,t)$ of the $(Cu_{50}Zr_{50})_{94}Al_4$ liquid in a semi-logarithmic representation, which shows that the strength of DH increases with cooling. The strength of $\chi_T(Q,t)$ indicates the extent of spatial correlation in the atomic motion[31] (Fig 1b). The strength of DH in $(Cu_{50}Zr_{50})_{100-x}Al_x$ liquids is quite similar, but slightly increases with increasing concentration of Al. These experimental results confirm the presence of dynamic heterogeneities in $(Cu_{50}Zr_{50})_{100-x}Al_x$ even well above the melting temperature and the increase in the length scale of correlated atomic motion in cooling the liquids.

The MD simulations were performed with a system of 100,000 atoms using the LAMMPS software and employing the potential developed by Sheng et al.[32] (see Supporting Information). To estimate the strength of DH and its temperature dependence, we first evaluated the self-part of the four-point correlation function[21],

$$\chi_4(t) = \frac{V}{k_B T N^2}[\langle Q_s^2(t) \rangle - \langle Q_s(t) \rangle^2], \tag{1}$$

where $Q_s(t) = \sum_{l=1}^{N} w(|r_l(t) - r_l(0)|)$ and $w(|r_1 - r_2|)$ are overlap functions that are unity for $|r_1 - r_2| \leq a$ and 0 otherwise (Fig. 1 c). We chose the distance parameter a = 1, which is the plateau value of the square of mean square displacement[21]. Although the absolute values of the strength of $\chi_T(Q,t)$ and $\chi_4(Q,t)$ obtained from the QENS and MD simulation cannot be compared, the growth rates with respect to the temperature are very similar (see Figs. 1b & 1c). Our results indicate that the relation between $\chi_T(Q,t)$ and $\chi_4(Q,t)$ proposed by Berthier and co-workers ($\chi_4(Q,t) \geq [k_B T^2/c_P]|\chi_T(Q,t)|^2$) holds good in these liquids.

To evaluate types of LGOs in these alloy liquids at different temperatures, we carry out Voronoi tessellation analysis of the atomic configurations generated by the MD simulations. The population of the five most abundant Voronoi clusters in the undercooled state (923 K) of the alloy liquids are shown in Fig. 2a. There are several types of polyhedrons in these metallic liquids at all temperature ranges[33]. However, the most abundant polyhedron in all four systems, above the melting temperature, was found to be <0,3,6,4> polyhedron, while the next abundant polyhedron was <0,2,8,2>. The population of these two polyhedrons increases with the Al concentration, while the growth of a specific polyhedron in cooling the liquids is dependent on the type of polyhedrons. The growth of the <0,3,6,4> polyhedron, which is most abundant in these liquids, is saturated in the supercooled liquid (see Fig. 2b). The next most abundant <0,2,8,2> polyhedron has grown almost linearly while cooling the liquids (see Fig. 2c). Interestingly, the icosahedra, <0,12,0,0> has grown much faster in the undercooled liquids (see Fig. 2d). In cooling these four alloy liquids, the growth of specific polyhedrons is similar, but the percentage of icosahedrons increased with the concentration of Al at any given temperature (Fig. 2d).

The dynamic-length-scale in these alloy liquids at various temperatures was obtained by the following procedures. First, we calculated the four-point, time-dependent structure factor for self-overlapping particles, $S_4(q,t)$, which is defined as,

$$S_4(q,t) = \frac{L^2}{N^2} \langle \rho(q,t)\rho(-q,t) \rangle \qquad (2)$$

where $\rho(q,t) = \sum_{j=1}^{N} w(|r_j(t) - r_j(0)|) \exp[iq \cdot r_j(0)]$, and $t$ is the time at which the maximum of dynamical four-point susceptibility $\chi_4(t)$ [18,34,35]. The DH is transient in time, reaching a maximum value at around the structural relaxation time, which measures the degree of cooperativity of structural relaxation. Second, the $S_4$ obtained at low-q values at a given temperature were fitted with Ornstein-Zernike form (see Fig 3b),

$$S_4(q,t) = \frac{S_4(q \to 0, \tau_4)}{[1+(q\xi_4)^2]} \qquad (3)$$

where $\xi_4$ is the dynamic correlation length or dynamic-length-scale. In decreasing the temperature, the $\xi_4$ increases exponentially (see Fig 3b). However, the growth rate of $\xi_4$ depends on alloy composition, with a higher amount of the Al content in the liquids a faster growth rate of $\xi_4$ was observed. In the $(Cu_{50}Zr_{50})Al_{10}$, which has the highest Al content, the $\xi_4$ increasing from 2 to 7.5, while in alloys with 2% Al, $\xi_4$ increases marginally (see Fig. 3b). As we compare the growth of $\xi_4$ with different polyhedrons, the growth rate shows two regimes for all types of polyhedron other than the icosahedron. Above the melting temperature ($T_m$) of the alloys, $\xi_4$ grows to some degree with <0,3,6,4> or <0,2,8,2> polyhedrons, but increases substantially below $T_m$. Surprisingly, the growth of $\xi_4$ shows a direct correlation with the population of icosahedrons in the liquids. This relation holds good in these alloy liquids in a temperature range as low as 300 K below $T_m$ and as high as 300 K above $T_m$.

Our study experimentally confirmed the existence of the dynamic heterogeneity in the glass-forming $(Cu_{50}Zr_{50})_{100-x}Al_x$ liquids and the corresponding length scale of correlated motion over a temperature range of 600 K. This is visible from the stretching of intermediate scattering function and the strength of dynamic susceptibility, respectively. Both the dynamic heterogeneity and the dynamic length scale were found to increase with Al content and cooling of alloy liquids. At the same time, the atomic mobility was found to decrease, which indicates strong coupling between structural relaxation and the dynamic-length-scale. Using the MD simulation, we quantitatively determined the population of different polyhedrons as a function of temperature in these liquids. Although the population of the icosahedron is very little in high temperature liquids, it increases substantially in an undercooled state. The population of icosahedrons grew exponentially in the range of temperature studied. Similarly, the dynamic length scale evaluated from the four-point dynamic structure factor increases exponentially with a decreasing in temperature. However, the growth of other polyhedrons did not follow specific temperature dependence. In fact, the growth of most abundant <0,3,6,4> polyhedron saturated in undercooled liquids, and the <0,2,8,2> polyhedron grows almost linearly with temperature. This suggests a strong correlation between the increase of the dynamic-length-scale and growth of icosahedrons in liquids. Additionally, as a function of Al content, the number of icosahedra increases and atomic dynamics slows down. This result suggests that the slowdown of the atomic dynamic in these liquids must have a strong coupling to the growth of icosahedron.

Entire existing studies in which both static and dynamic-length-scale have been computed for the same glass-forming liquids show that these length scales do not correlated each other[18,5,6]. The value of static-length-scales calculated in various studies generally found to be smaller than the dynamic one and the difference increases with decrease in temperature[36,37]. Therefore, we did not attempt to calculate the static-length-scale in the $(Cu_{50}Zr_{50})_{100-x}Al_x$ liquids. However, dynamic-length-scales obtained in our study exhibit a power law behaviour with relaxation time; ($\tau_\alpha = a\xi_4^z$), where $a$ is constant and $z$ is the dynamic critical exponent. The value of $z$ varied from 4.37 – 10.06 (see Fig. 3f). Such a power law behaviour has been reported in many glass-forimg liquids [21,38]. It has also been shown that the icosahedral clusters have a strong tendency for connectivity and form a string-like network[39–41]. Therefore, we explored the possibility of a correlation between the increase of dynamic-length-scale and growth of icosahedra in these alloy liquids. We scaled the population of the icosahedron with dynamic-length-scale in the four alloy liquids investigated. Surprisingly, a linear relationship was observed in the alloy liquids (Fig. 3e) investigated. To our knowledge, such a result has never been shown before. Our result provides evidence that the transient icosahedral connectivity could be the origin of the dynamic length scale in glass-forming liquids.

In summary, we have studied the growth of dynamic-length-scale and local-geometrical orders in the glass-forming $(Cu_{50}Zr_{50})_{100-x}Al_x$ liquids over 600 K using QENS and MD simulation. The intermediate scattering functions obtained from the QENS data indicate that the heterogeneous dynamics in $(Cu_{50}Zr_{50})_{100-x}Al_x$ liquids and MD simulation results proves the existence of such dynamics. The dynamic-length-scale has been evaluated by the MD simulation at various temperatures in $(Cu_{50}Zr_{50})_{100-x}Al_x$ liquids by the four-point, time-dependent structure factor. Further, using the MD generated atomic configurations, we have evaluated different local-geometrical-orders present in these liquids at various temperatures and compared with the dynamic-length-scale. We have observed that, although various local-geometrical orders are present in the liquids, the increasing number of icosahedron shows a linear variation with the growth of dynamic-length-scale. Our result shows for the first time that the transient icosahedral ordering could be the origin of the dynamic-length-scale in glass-forming liquids.

**ASSOCIATED CONTENT**

**Supporting Information**

**Sample preparation:** The alloy samples for the quasielastic neutron scattering (QENS) were prepared by arc melting of the pure elements; Cu (99.99%), Zr (99.2%) and Al (99.999%). The arc melting were done in Ti-gettered argone atmosphere and samples were remelted several time to ensure homogeneous composition. The sample ingots were then powdered and filled in $Al_2O_3$ hollow cylindrical crucibles. The $Al_2O_3$ crucibles with powder samples were then remelted two-three times in a very high vacuum furnace. The crucible provides a hollow cylindrical geometry to the samples with a 22mm diameter, 30mm height and 1.2mm thickness. The small thickness is to avoid multiple scattering and with the sample thickness of 1.5 mm only about 8% incoming neutron will be scattered.

**Quasielastic neutron scattering experiments:** The quasi-elastic neutron scattering (QENS) experiment was performed on the time-of-flight neutron spectrometer, Pelican, at the OPAL reactor, Australian Nuclear Science and Technology Organisation, Australia. The wavelength of the incident neutron was 6 Å which provides a Q range of 0.2 Å$^{-1}$ – 1.9 Å$^{-1}$ and an energy resolution of 70 µeV for the experimental setup. Since the maximum of static structure factors of the alloy melts are around 2.8 Å$^{-1}$[42], the scattered signals from our experiments are mainly due to the incoherent scattering from the Cu atoms of the samples. Therefore, the dynamics that we observed from the QENS data are of the self-dynamics of Cu atoms in the respective alloy liquids. For the high temperature measurements, we used an ILL-design vacuum furnace. The QENS data were collected from high temperature down to the melting temperature of each alloy (1573 K, 1473 K, 1373 K, 1273 K, and 1223 K). The QENS data were collected over a duration of 4 hours at each temperature. An empty $Al_2O_3$ sample holder was also measured at each temperature for background subtraction and self-absorption correction. For correcting the detector efficiency, a vanadium crucible with a similar sample geometry was also measured at room temperature for 4 hours.

**QENS data analysis:** The raw data was normalized to the vanadium data, and converted to the dynamic structure factor using LAMP (Large Array Manipulation Program). The self-absorption of the sample S in a container C was corrected using

$$S_S(2\theta,\omega) = \frac{1}{A_{S,SC}(2\theta,\omega)} S_{S+C}(2\theta,\omega) - \frac{A_{C,SC}(2\theta,\omega)}{A_{S,SC}(2\theta,\omega)A_{C,C}(2\theta,\omega)} S_C(2\theta,\omega),$$

where $S_C$ and $S_{S+C}$ is the scattering from the container, and from both sample and container respectively. $A_{C,C}(2\theta,\omega)$ is the correction factor for scattering and self-absorption of the container, $A_{C,SC}(2\theta,\omega)$ is the correction factor for scattering due to the container and absorption in both sample and container and $A_{S,SC}(2\theta,\omega)$ is the correction factor for scattering due to the sample and absorption in both sample and container. At last, the intermediate scattering functions Φ(Q,t) were obtained by Fourier transforming the dynamic structure factor.

**Molecular Dynamics Simulation:** The molecular dynamics (MD) simulations were done for the same compositions measured in QENS experiments using Large-scale Atomic/Molecular Massively Parallel Simulator (LAMMPS); a free software obtained from the Sandia National Laboratories, USA. The embedded-atom method (EAM) potential was used to describe the interatomic interactions. The time step used in the simulation was 2 fs and periodic boundary conditions were applied. The Nose-Hoover thermostat was used to control the temperature. For each system, the initial configuration containing 100,000 atoms was equilibrated at 2000 K for 5 ns followed by rapid quenching to the desired temperatures with the rate of $2 \times 10^{11}\text{Ks}^{-1}$ in NPT ensemble. The volume of the system was adjusted to give zero pressure during cooling. Before taking the structural configurations, the systems were relaxed for extra 1 ns by the NVT ensemble. Voronoi tessellation calculates the polyhedral cells which have planar faces and completely fill the space by constructing bisecting planes along the lines connecting the target atom and its neighbors. The polyhedrons surrounding the central atom are described by the Voronoi index <$n_3,n_4,n_5,n_6$…>, where $n_i$ is the number of i-edged faces of the polyhedron. A cutoff distance of 5Å was used in this study.

**AUTHOR INFORMATION**

**Correspondence.**

Suresh M. Chathoth, Email: smavilac@cityu.edu.hk

**Acknowledgements**

This research has been supported by City University Strategic Research Grant, No: 7004695. The authors thank ANSTO for providing beam time on the Pelican instrument at the OPAL reactor.

**Competing financial interests**

The authors declare that they have no competing financial interests.

**Figures and Captions**

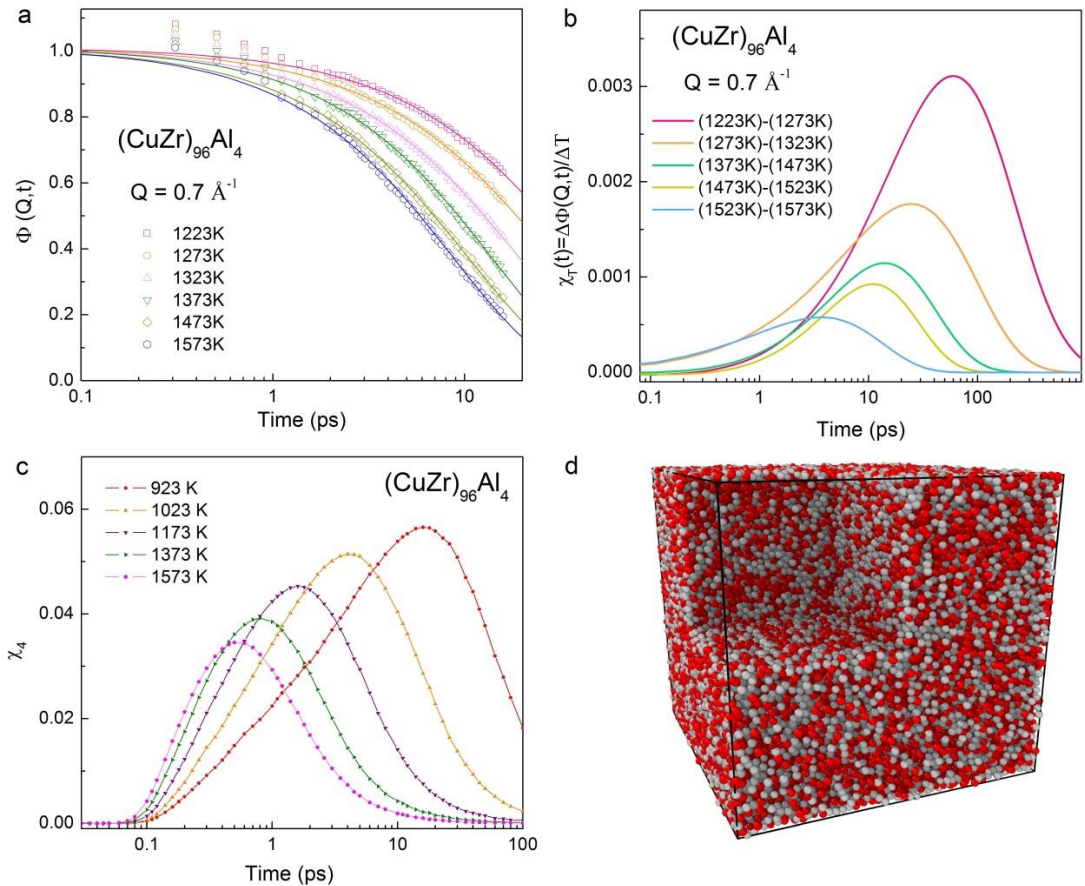

**Figure 1** (a) The intermediate scattering function $\Phi(Q,t)$ of $(CuZr)_{96}Al_4$ at a Q value of 0.7 $Å^{-1}$ at six measured temperatures. The solid lines are the fitting of the KWW function in the α-relaxation time range. (b) The dynamic susceptibility $\chi_T(t)$ of $(CuZr)_{96}Al_4$ at a Q value of 0.7 Å$^{-1}$ on a linear-log scale. (c) The four-point correlation function $\chi_4$ for all types of atoms of (Cu$_{50}$Zr$_{50}$)Al$_4$ liquid. (d) A snapshot of (Cu$_{50}$Zr$_{50}$)Al$_4$ liquid at 923 K with mobility of grey atoms substantially slower than red atoms.

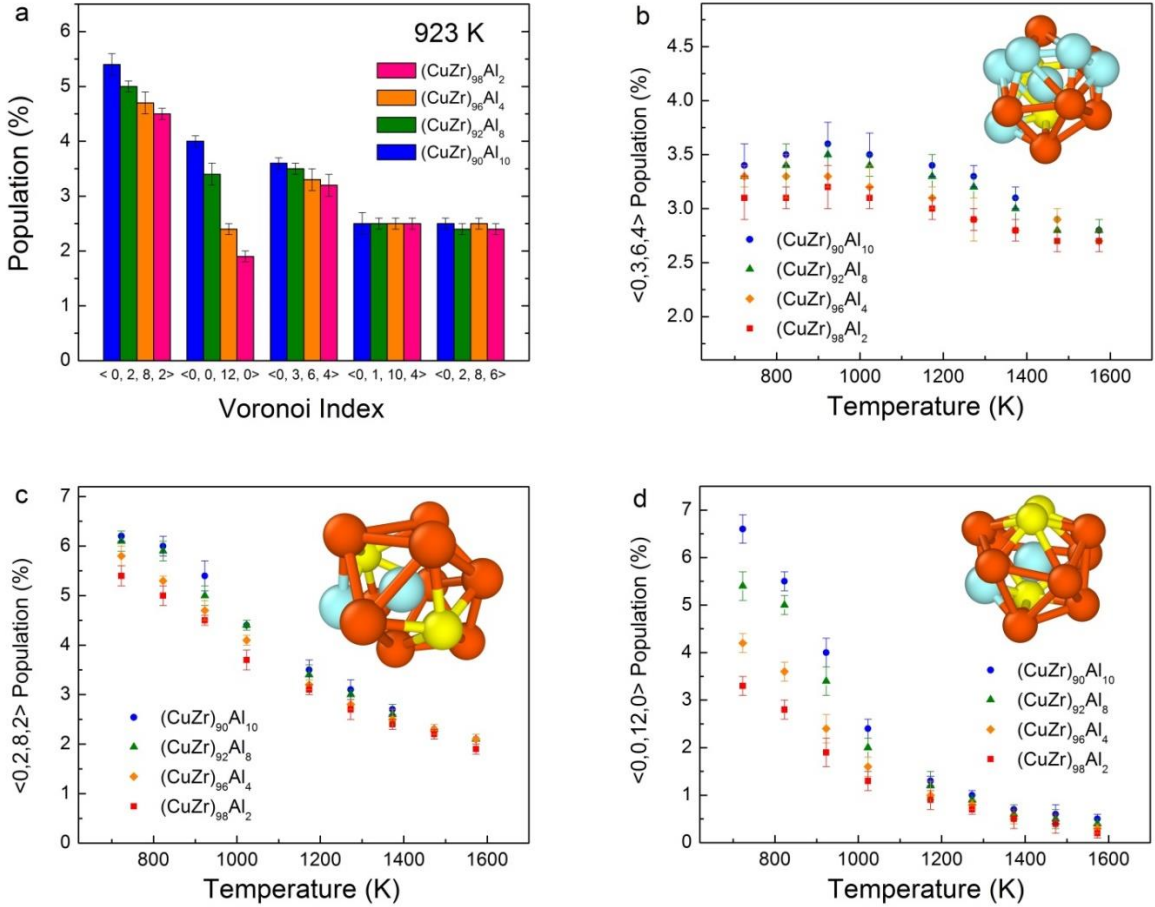

**Figure 2**: (a) The population of the most five dominant polyhedrons for the four systems in an undercooled temperature (923 K). The polyhedron population of (b) <0,3,6,4> (c) <0,2,8,2>, and (d) <0,0,12,0> as function of temperature. Inserts represent the respective polyhedrons.

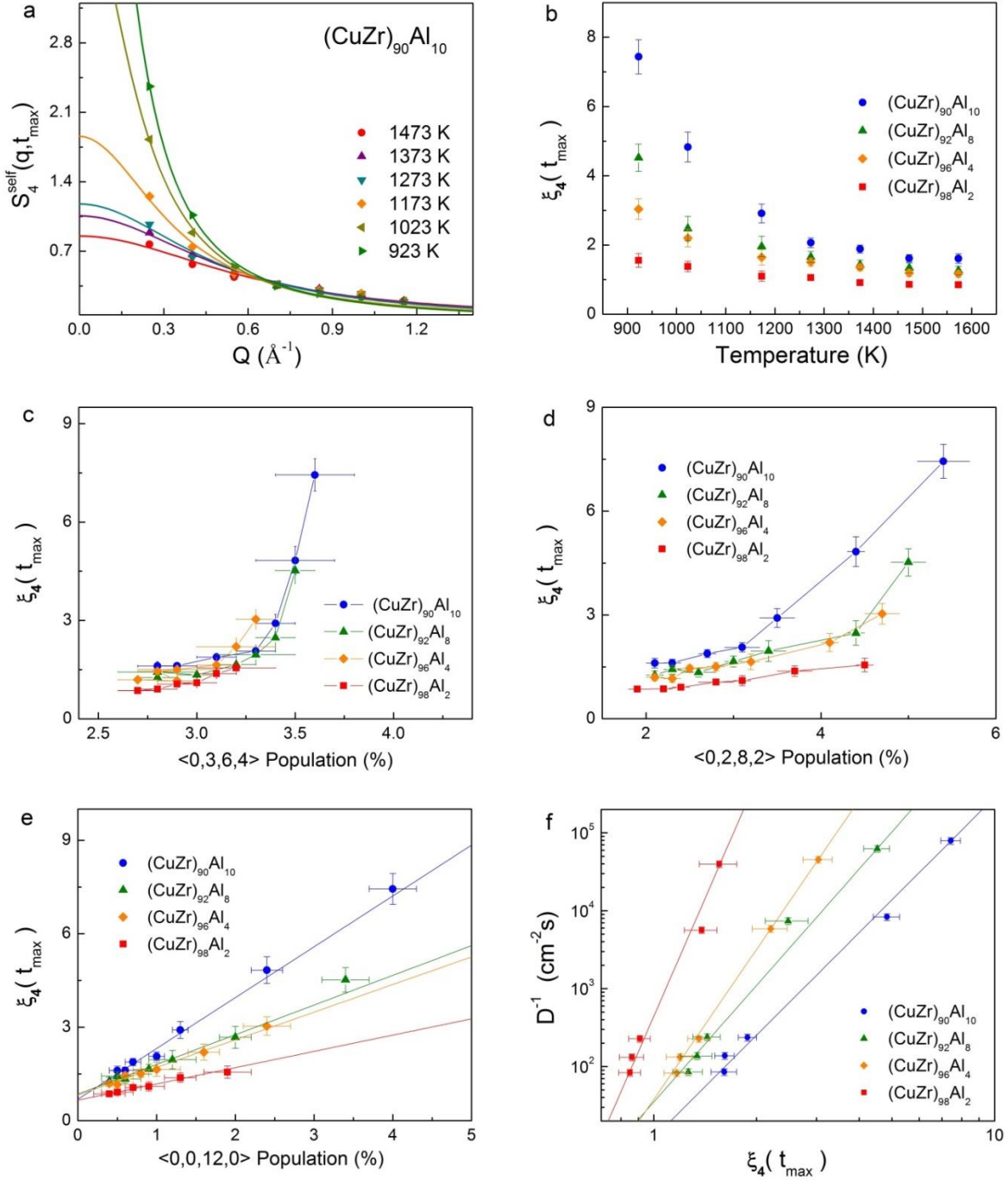

**Figure 3:** (a) The four-point dynamic structure factor of $(Cu_{50}Zr_{50})Al_{10}$ at various temperatures, while solid lines fit the Ornstein-Zernicke form (Eq. 3). (b) The dynamic correlation length or the dynamic-length-scale ($\xi_4$) at various temperatures for the alloy liquids studied. The growth of $\xi_4$ as a function of (c) <0,3,6,4>, (d) <0,2,8,2>, and (e) <0,0,12,0> polyhedrons. The inverse of diffusion (relaxation time) shows a power law behaviour with the dynamic correlation length. The straight lines represent how the power law fits ($D^{-1} = a\xi_4^z$).